# High $T_c$ Hydrides: Interplay of Optical and Acoustic Modes and Comments Regarding the Upper Limit of $T_c$

Vladimir Z.Kresin

Lawrence Berkeley Laboratory, University of California at Berkeley,CA 94720

Abstract.

Phonon spectrum of many superconducting compounds and, especially, high $T_c$ hydrides, is broad and rather complicated ,because of presence of high frequency optical modes. In order to analyze an interplay of optical and acoustic phonon branches,it is convenient to introduce two coupling constants, $\lambda_{opt.}$ and $\lambda_{ac.}$, along with to characteristic frequencies. The correlation between the value of $T_c$ with this interplay is demonstrated for the family of tantal hydrides ($TaH_2$/ $TaH_4$/ $TaH_6$). The problem of upper limit of $T_c$ is discussed. The phenomenon of room temperature superconductivity can be provided by the electron-phonon interaction and described by the strong coupling theory.

Recent milestone discovery of the sulfur hydride with record value of $T_c \approx 203K$ under high pressure [1,2] has resulted in renewed interest to the phonon mechanism of pairing. This paper focuses on peculiar aspects of this mechanism in the hydrides. The pioneering theoretical papers [3,4] and studies [5,6] on sulfur hydrides demonstrated that phonon mechanism can provide record values of the critical temperature. However, we know that the phonon spectrum in the hydrides is very broad (up to 200-250 meV) and has a rather complicated structure. That's why the evaluation of the critical temperature should be performed with a special care. This is especially important ,because it is perfectly realistic that not only sulfur hydrides, but other hydrides can display high values of $T_c$. In connection with this, a number of theoretical papers containing predictions of high values of Tc, have been published (see, e.g., reviews [7-9]). Generally speaking, high values of $T_c$ can be provided by large values of the characteristic phonon frequency and the coupling constant. Nevertheless, the structure of the phonon spectrum in the hydrides is very complex, so that the usual expressions for the critical temperature (see, e.g.,[10-14]) should be generalized (see below).As a result, it is possible to analyze the interplay of different phonon branches ( optical vs. acoustic) for various families of hydrides.

*Main relations*. The value of the critical temperature for the superconducting state caused by the electron-phonon interaction can be determined from the following equation ([ 1 ],see review [16]):

$$\Delta(\omega_n)Z = \pi T \sum_m \int d\Omega \frac{\alpha^2(\Omega)F(\Omega)}{\Omega} D(\Omega, \omega_n - \omega_m) \frac{\Delta(\omega_m)}{[\omega_m^2 + \Delta^2(\omega_m)]^{1/2}}$$

(1)

here Z is the renormalization function describing usual electron-phonon scattering:

$$Z = 1 + \pi T \sum_m \int d\Omega \frac{\alpha^2(\Omega)F(\Omega)}{\Omega} D(\Omega, \omega_n - \omega_m) \frac{\omega_m}{[\omega_m^2 + \Delta^2(\omega_m)]^{1/2}}$$

(1')

In Eqs.( 1 ),( 1' ) $\omega_n = (2n+1)\pi T$ , $\Delta(\omega_n)$ is the pairing order parameter,

$$D = \frac{\Omega^2}{\Omega^2 + (\omega_n - \omega_m)^2} \qquad (2)$$

is the phonon propagator, and

$$<F^+> = \frac{\Delta(\omega_m)}{\left[\omega_m^2 + \Delta^2(\omega_m)\right]^{1/2}} \qquad (2')$$

$F^+$ is the pairing function [17]. We employed the method of thermodynamic Green's functions (see, e.g., [18]) The important ingredient of the theory is the spectral function S($\Omega$)= $\alpha^2(\Omega)F(\Omega)$ ; here F($\Omega$) is the phonon density of states, and $\alpha^2(\Omega)$ contains the electron-phonon matrix element (see review [19]). This function and the Coulomb pseudopotential $\mu^*$ can be determined with use of special tunneling spectroscopy (see [20] and the reviews [21-23]).In the absence of the tunneling data, one can use this function obtained from ab initio calculations. As for the coupling constant $\lambda$ ,it also determined from the function S($\Omega$) :

$$\lambda = 2\int d\Omega \frac{\alpha^2(\Omega)F(\Omega)}{\Omega} \qquad (3)$$

Let us stress ,at first, that the Eq.(1 ) does not contain explicitly the electron-phonon coupling constant. The value of $T_c$ can be evaluated directly from Eq.(1) without invoking the concept of coupling constant (see,e.g.,[24-27] ). At the same time this concept is very useful, since it allows to obtain analytical expressions for $T_c$ and study various dependences. However, and this can be seen directly from Eq.( 1),to introduce the coupling constant $\lambda$ is not a trivial step, because the phonon frequency enters not only into the function $\alpha^2(\Omega)F(\Omega)$ , but also into the phonon propagator D. The situation greatly simplified in the weak coupling case when $T_c \ll \Omega_{ch.} \approx \Omega_D$ ; then $D \approx 1$. The coupling constant, defined by Eq.(3) can be factored out; as a result, we obtain usual logarithmic BCS dependence [28]. For the intermediate and strong electron-phonon coupling the phonon propagator cannot be neglected and its dependence on the phonon frequency is an important ingredient in the integrand.

Usually the phonon propagator D, defined by Eq.(2) is taken at some characteristic frequency $\tilde{\Omega}$, so that $D = \tilde{\Omega}^2 \left[ \tilde{\Omega}^2 + (\omega_n - \omega_{n'})^2 \right]^{-1}$. Then Eq. (1) can be written in the form (at $T=T_c$):

$$\Delta(\omega_n)Z = \pi T_c \lambda \sum_m D(\tilde{\Omega}, \omega_n - \omega_m) \frac{\Delta(\omega_m)}{|\omega_m|} \Big|_{T_c}$$

(4)

$$Z = \pi T_c \lambda \sum_m D(\tilde{\Omega}, \omega_n - \omega_m) \frac{\omega_m}{|\omega_m|} \Big|_{T_c}$$

(4')

The coupling constant is defined by Eq.( 3 ). One may think that the replacement $D(\Omega, \omega_n - \omega_{n'}) = D(\tilde{\Omega}, \omega_n - \omega_{n'})$ just corresponds to the trivial step like

$$\int f(x)g(x)dx = g(\tilde{x}) \int f(x)dx$$

. However, this is not the case, because the D-function depends not only on , but also on $\omega_n - \omega_{n'}$ with summation over n'. Because of this factor, the average $\tilde{\Omega}$ , strictly speaking, depends on the value of $n'$ and is different for various terms.

In reality, the replacement $\Omega \to \tilde{\Omega}$ is an approximation and its use is based on the difference between the characteristic scales for the phonon density of states $F(\Omega)$ and the phonon propagator D. If we are dealing with usual superconductors and their acoustic phonon branches (longitudinal and transverse), then the phonon density of states contains two sharp and relatively close peaks, corresponding to the short wave regions, where the phonon dispersion is almost flat (where the density of states $F(\Omega) \propto q^2 dq/d\Omega$ is large). The peak region corresponds to relatively narrow interval of phonon frequencies; the phonon propagator is almost constant within this interval and it allows to replace $\Omega$ by some characteristic value of $\tilde{\Omega}$. This value is taken sometimes as $\tilde{\Omega}_{\log} = (2/\lambda) \int d\Omega \alpha^2(\Omega) F(\Omega) \log \Omega$ (see. e.g., [12, 29,30]), although, from general principles, under the "log" sign there should be always the dimensionless quantity. Another generally accepted quantity is

$$\tilde{\Omega} = <\Omega^2>^{1/2} = (2/\lambda) \int d\Omega \alpha^2(\Omega) F(\Omega) \Omega$$. This quantity is a good

approximation for $\tilde{\Omega}$ (see [31]) and will be used below.

However, for the new high Tc family, superconducting hydrides, the situation is more complicated. The presence of two ions in unit cell, including light hydrogen ion, leads to an appearance ( in addition to usual acoustic branches) of high frequency optical modes. As a result, the phonon spectrum is very broad and the phonon density of states contains many peaks (see Fig.1).For such a system, the description based on introduction of single characteristic frequency and, correspondingly, single coupling constant ,as in Eq.( 4), is too crude. In order to analyze the interplay of different phonon modes, it is convenient to separate the contributions of the optical and acoustic branches and, correspondingly, to introduce [31,7] two characteristic phonon frequencies $\tilde{\Omega}_{opt.}$, $\tilde{\Omega}_{ac.}$ and two coupling constants $\lambda_{opt.}$, $\lambda_{ac.}$, so that

$$\lambda_i = \int_i d\Omega \alpha^2(\Omega)F(\Omega)/\Omega; \tilde{\Omega}_i = <\Omega^2>_i^{1/2}$$

(5)

$$<\Omega^2>_i = (2/\lambda_i)\int_i d\Omega\Omega\alpha^2(\Omega)F(\Omega)$$

i ≡ {opt., ac.}.

The equation determining the value of $T_c$ has a form (cf. Eq. (4)):

$$\Delta(\omega_n)Z = \pi T_c \sum_m \int d\Omega[(\lambda_{opt.} - \mu^*)D(\tilde{\Omega}_{opt}, \omega_n - \omega_m) + \lambda_{ac.}D(\tilde{\Omega}_{ac}, \omega_n - \omega_m)]\frac{\Delta(\omega_m)}{|\omega_m|}$$

(6)

If the dominant role is played by optical phonons, the solution of Eq.( 6 ) is [31] :

$$T_c = \left[1 + 2\frac{\lambda_{ac}}{\lambda_{opt} - \mu^*}\frac{1}{1+\left(\pi T_c^0/\tilde{\Omega}_{ac}\right)^2}\right]T_c^0$$

(7)

where $T_c^0$ is determined by contribution of optical phonons only. For $T_c^0$ one can use the McMillan-Dynes expression ($\lambda_{opt.} \lesssim 1.5$):

$$T_c^0 = \left(\tilde{\Omega}_{opt}/1.2\right)\exp\left[\frac{1.04(1+\lambda_{opt.})}{\lambda_{opt.} - \mu^*\left(1 + 0.62\lambda_{opt.}\right)}\right]$$

( 8 )

or the expression obtained analytically in [13 ],which is close to Eq.( 8 ).For larger values of $\lambda_{opt.}$ one can use the obtained by Allen- Dynes expression [12 ] ,or the expression obtained by the author [14 ].

The values of the coupling constants and the frequencies $\tilde{\Omega}_{opt.}$ and $\tilde{\Omega}_{ac.}$ can be determined from the spectral function $\alpha^2(\Omega)F(\Omega)$. As was mentioned above, the special method of tunneling spectroscopy allows us to obtain this function. In the absence of such data one can use the ab initio calculations. This approach was used by Gor'kov and the author in [ 32 ] (see also review [ 7 ]) to describe the superconducting states in the *R3m* ($T_c \approx 100K$) and *Im-3m* ($T_c \approx 203K$) phases of sulfur hydride. We used the ab initio calculations for these phases carried out in [6].

*Tantal hydrides*. The approach, described above, was used in [ 32 ],[ 7] to evaluate the values of $T_c$ and the isotope coefficients for the sulfur hydride with various phases including that with the record value of $T_c$, and the $CaH_6$ system, studied in [ 33 ]. In this section we consider different example, namely, the Ta-H system. This compound was described recently in theoretical paper [ 34]. It is important that the authors [34] evaluated the spectral functions for three different compounds: $TaH_2$, $TaH_4$, $TaH_6$. The critical temperatures were also calculated in [ 34 ] with use of the Allen-Dynes equation. Below we consider these compounds with the two-coupling constants model. In addition to an increased accuracy it will allows us to analyze factors, controlling the values of $T_c$ and the isotope coefficients (see below, Table I). The possibility to analyze a whole family of Ta-H compounds allows us to demonstrate the the physical picture, which can be provided by the two-coupling constants method.

Let us start with the hydride $TaH_6$. The spectral function is presented on Fig. 1a. Let us split the phonon spectrum into two regions: region I ($\Omega < 15$ THZ) and region II ($\Omega > 15$ THZ). With use of Eqs.( 5 ) one can calculate the characteristic frequencies, and we obtain: $\tilde{\Omega}_{opt.} \approx 1440K$, and $\tilde{\Omega}_{ac.} \approx 200K$. We assume here and below that $\mu^* = 0.1$. The curve (Fig. 1 ) allows to determine the values of the coupling constants.

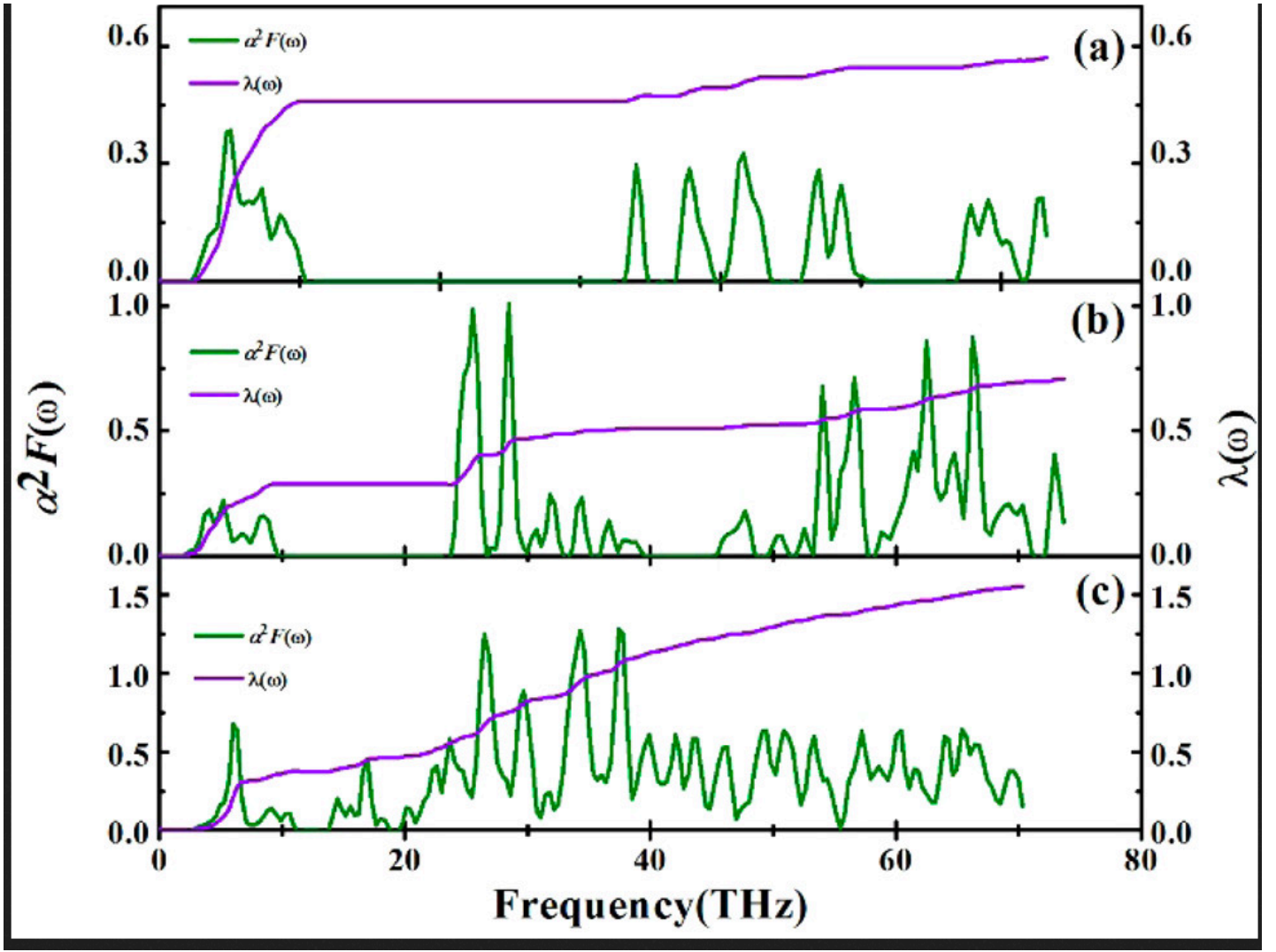

Fig.1. The spectral function $\alpha^2(\Omega) F(\Omega)$ for: a) $TaH_2$; b) $TaH_4$; c) $TaH_6$

One can see, that $\lambda_{opt.} \approx 1.1$ and $\lambda_{ac.} \approx 0.4$. With use of Eqs. (7) ,one can calculate the value of $T_c$ ,which appears to be equal to $T_c \approx 140K$ for the $TaH_6$ compound. Note that $T_c^0 \approx 115K; \Delta T_{c;ac.} \approx 25K$ .

As the next step, one can evaluate the value of the isotope coefficient . Since the optical modes corresponds, mainly, to the hydrogen ions motion, whereas the acoustic modes describe the motion of heavy Ta ions, the value of the isotope coefficient for the H→D substitution reflects the relative contribution of the high frequency optical modes. One can use the expression [31],[ 7 ]:

$$\alpha \approx \frac{1}{2}\left[1 - 4\frac{\lambda_{ac}}{\lambda_{opt}}\frac{\rho^2}{(\rho^2+1)^2}\right] \quad (9)$$

We obtain for TaH$_6$ the value $\alpha \approx 0.4$. It is close to the limiting value = 0.5 and it means that the optical modes are dominant for the pairing in the TaH$_6$ system. The values of T$_c$ and $\alpha$ are consistent with the assumption that Eq.( 7 ) is valid, if the optical modes are dominant.

Let us turn our attention to the TaH$_4$ system. Similarly to that performed above for the TaH$_6$ compound, one can evaluate $\tilde{\Omega}_{opt.} \approx 2200K$, and $\tilde{\Omega}_{ac.} \approx 265K$ , and also $\lambda_{opt.} \approx 0.35$ , and $\lambda_{ac.} \approx 0.35$ . One can see that the values of the coupling constants are almost equal, and for such a case the Eqs ( 7 ) , ( 9 ) are not applicable. However, in this case the strengths of the coupling are relatively small, and one can use the usual logarithmic approximation .One should take into account the presence of two peaks at $\tilde{\Omega}_{opt.}$ and $\tilde{\Omega}_{ac.}$ (cf.,the analysis of superconductivity for the *R3M* structure ( T$_c \approx$100K ) in sulfur hydrides in [ 31 ],[ 7 ]). Then we obtain (see the evaluation of the pre-exponential factor in [14,35 ] ):

$$T_c \approx 0.25\left(\tilde{\Omega}_{opt}\right)^{\frac{\lambda_{opt}}{\lambda_T}} \left(\tilde{\Omega}_{ac}\right)^{\frac{\lambda_{ac}}{\lambda_T}} \exp\left[-\frac{1}{\lambda_T - \mu^*}\right] \quad (10)$$

($\lambda_T = \lambda_{opt.} + \lambda_{ac.}$), and T$_c \approx$36K .

The isotope coefficient can be evaluated from Eq.(10 ) and the relation (see [7] ): $\alpha = 0.5\left(\tilde{\Omega}_{opt.}/T\right)\left(\partial T_c / \partial \tilde{\Omega}_{opt.}\right)$. Then we obtain: $\alpha = 0.5(\lambda_{opt}/\lambda_T)$ and for the TaH$_4$ its value is $\alpha \approx 0.25$.

Finally, for the TaH$_2$ hydride one can determine from Fig. 1 that $\tilde{\Omega}_{opt.} \approx 2600K$, and $\tilde{\Omega}_{ac.} \approx 70K$. As for the coupling constants, we obtain $\lambda_{opt.} \approx 0.1$, and $\lambda_{ac.} \approx 0.45$. Then $T_c$ can be determined for Eq.(10) and is equal to $T_c \approx 3.5K$. The isotope coefficient for the H→D substitution appears to be small: $\alpha \approx 0.1$.

The discovery of new family of high $T_c$ superconductors, high $T_c$ hydrides, with record value of $T_c$, made by M.Eremets and his collaborators, was described in [1,2]. As was stressed above, a number of other hydrides with high values of the critical temperature, up to room temperature, have been predicted, e.g., the families of La-H and Y-H compounds (see,e.g.,[8],[36]) and the reviews {7-9})

One can see from the Table I that an increase in the value of the critical temperature for the sequence TaH$_2$ → TaH$_4$ → TaH$_6$ is caused by redistribution of the electron-phonon interaction between the optical and acoustic modes. This redistribution is also reflected in the values of the isotope coefficient for the H→D substitution. This value is growing with an in increase in $T_c$. The strength of the $\lambda_{opt}$ is directly related to the increase in a number of hydrogen ions modes, which is growing with an increase in a number of H ions in the unit cell. As we can see, the interplay of the introduced coupling constants is directly related to behavior of $T_c$.

Table I

| Hydride | $\lambda_{total}$ | $\lambda_{opt.}$ | $\lambda_{ac.}$ | $\lambda_o / \lambda_a$ | $T_c$ | $\alpha$ |
|---|---|---|---|---|---|---|
| TaH$_2$ | 0.55 | 0.1 | 0.45 | 0.22 | 3.5K | 0.1 |
| TaH$_4$ | 0.7 | 0.35 | 0.35 | 1 | 36K | 0.25 |
| TaH$_6$ | 1.5 | 1.1 | 0.4 | 2.75 | 140K | 0.4 |

*About stability of lattice and upper limit of $T_c$* According to recent paper [37],the value of Tc is restricted by the inequality $T_c < a\tilde{\Omega}$, $a \approx 0.1$. This conclusion is supported by empirical observations along with some Monte Carlo calculations [38]. The authors use the Debay frequency as the characteristic phonon frequency; in reality different value of $\tilde{\Omega}$ enters the theory (see above, Eq.(5)),which is much below $\Omega_D$ (for example, for Pb $\tilde{\Omega} \approx 5.5$ meV, whereas $\Omega_D$ is almost twice larger [23] .Let us note, at first, that, as was stressed above, in the presence of high frequency optical phonon modes the description with use of single characteristic phonon frequency is a rather crude approximation. If we introduce two characteristic frequencies (see above) ,then the criterion [36] becomes uncertain. Note also that for sulfur hydride (see, e.g., [24]). the value of the single characteristic frequency is of order of $\tilde{\Omega} \approx 1500$K,whereas $T_c \approx 230$K , so that  the criterion formulated in [37 ] is violated. Another example if for yttrium hydrides  [8 ] : $T_c \approx (305-326)$K   for $YH_{10}$  (P=250GPa),whereas $\tilde{\Omega} \approx 1300$K.

The authors [ 37 ] also stated that the existing theory for the intermediate and strong coupling superconductivity (sometimes it is called the Migdal-Eliashberg [39,15] theory (ME)) is not valid at $\lambda>1$,because of bipolaron instability. This statement is not new and was formulated in [ 40 ]. However, it is essential to stress that for majority of real systems this statement is irrelevant. As we know ,there are a number of superconductors which are characterized by large values of $\lambda$ and are  described very well by the ME theory . Among them (see, e.g. [ 23 ]) Am-$Pb_{0.45}Bi_{0.55}$   compound ( $\lambda$= 2.6),  Pb ( $\lambda$ =1.4). New high $T_c$ hydrides are also characterized by large values of the coupling constant. For example, the superconducting state of the $H_3S$ compound with record value of the critical temperature ($T_c$= 203K) is provided by the coupling strengths: $\lambda_{opt.} \approx 1.5$, $\lambda_{ac.} \approx 0.5$,so that $\lambda_{tot.} \approx 2$. The large value of  the coupling strength was obtained in a number of papers (see, e.g., [ 8 ],and the reviews [ 7,9 ]).

The problem of lattice instability has an interesting history and we are not going to describe it here (see, e.g., [16]). At first, it was discussed in [39], where the limiting value of λ was introduced. However, the conclusion was based on the so-called Froehlich Hamiltonian, containing the measured phonon frequency and the electron-phonon interaction (EPI). As was analyzed in [41] and, especially in [42], this Hamiltonian does not follow from the usual adiabatic approximation (then $E_f >> \tilde{\Omega}$ [43], see, e.g., review [16]), since the formation of the phonon spectrum and EPI cannot be treated independently; EPI is the key ingredient forming the phonon spectrum. Therefore, the concept of instability [37,38] is not valid for the typical adiabatic case $E_F >> \tilde{\Omega}$. The Froehlich Hamiltonian is applicable for the anti-adiabatic case, when $\tilde{\Omega} >> E_F$; then the carrier concentration is relatively small and in this case we are dealing with the instability [39],[40], [37]. According to [37], the instability was obtained with use o the Holstein Hamiltonian by Monte Carlo simulations. This result is not surprising, since the Holstein Hamiltonian is similar to the Froehlich Hamiltonian. It was formulated by Holstein [44] for the semiconductors with small carrier concentration, that is, for the anti-adiabatic case.

As for the predictions of the limiting value of the critical temperature, the history of superconductivity knows several breakthroughs made contrary to the predictions. In reality, the lattice instability, probably, exists at some large values of the coupling constants, but this is still an open question. The recent discovery of high $T_c$ hydrides [1,2] with values of $T_c$ not distant from the room temperature clearly demonstrates that the strong coupling theory works for values of the coupling strength λ ≈2-3, sufficient for achievement of superconductivity at room temperature. We can look forward to this event, and there are all reasons to expect it in a near future.

*Acknowledgements.* The work is supported by the Lawrence Berkeley National Laboratory, University of California at Berkeley, and the U.S. Department of Energy.